# Evaluating Prediction Model Impact Using Observational Data: Clarifying the Implementation Pathway Before Estimation

Yingzi Yang[a,b], Xinyu Yang[a,b], Yuhao Li[a,b], Yuehan Duan[a,b], Xianglong Meng[a,b], Shengfeng Wang[a,b*], Junfeng Wang[c*]

**Affiliations:**

[a]Department of Epidemiology and Biostatistics, School of Public Health, Peking University, Beijing, China; [b]Key Laboratory of Epidemiology of Major Diseases (Peking University), Ministry of Education, Beijing, China; [c]Department of Epidemiology, CAPHRI Care and Public Health Research Institute, Maastricht University, Maastricht, Netherlands

***Correspondence:**

**Shengfeng Wang,** Department of Epidemiology and Biostatistics, School of Public Health, Peking University, Beijing, China, E-mail: shengfeng1984@126.com

**Junfeng Wang,** Department of Epidemiology, CAPHRI Care and Public Health Research Institute, Maastricht University, Maastricht, Netherlands. E-mail: junfeng.wang@maastrichtuniversity.nl

# Abstract

**Background:** Evaluating prediction model impact requires specifying how recommendations influence clinical decisions and treatment. Model use alone does not define implementation because clinicians may modify recommendations and intended treatments may not be executed. We developed a framework for specifying implementation strategies and their clinical consequences, particularly for observational evaluations.

**Methods:** We represented implementation as a sequence of model use, clinician decision making, treatment execution, and clinical outcomes. We defined target strategies, counterfactual expected outcomes, discrete contrasts, and a continuous implementation value surface. Three observational scenarios were distinguished by model implementation and observability of individual-level model use.

**Results:** We defined five strategies: no-model care and four model-guided strategies combining strict adherence to or clinician modification of model recommendations with full or constrained execution of intended treatment decisions. Their strategy-specific expected outcomes define implementation values. The value surface characterizes variation in adherence and execution within specified patient-level mechanisms. The three observational scenarios distinguish no implementation, implementation with model use recorded, and implementation with model use unrecorded, without conflating observed pathways with target strategies.

**Conclusions:** Prediction model impact evaluation requires explicit implementation strategies, not just an indicator of model use. The framework clarifies the clinical consequences to be compared and the implementation information available in observational data, providing a foundation for subsequent identification and estimation.

**Keywords**

Clinical prediction models; Impact evaluation; Observational data; Implementation strategies; Counterfactual outcomes; Clinician adherence; Treatment execution

**Key messages**

- Model use alone does not fully describe how prediction model recommendations affect treatment, because clinicians may modify recommendations and intended treatment decisions may not be executed.
- The framework represents implementation as a sequential process, defines five target strategies and their clinical outcome estimands, and distinguishes three observational scenarios by model implementation and the recording of individual-level model use.
- The framework distinguishes target strategies from observed care pathways, clarifying what impact evaluations compare and what implementation information observational data provide.

# 1 Introduction

Clinical prediction models increasingly support treatment decisions, but good discrimination and calibration do not establish that their use improves patient outcomes[1, 2]. Clinical impact depends on whether predictions influence decision making and patient management[3]. Evaluation must therefore consider the process through which recommendations become treatment.

Randomized impact studies directly assess model implementation but may be costly or infeasible[4]. Observational data offer an important alternative[5], and methods such as target trial emulation provide established approaches for evaluating treatment strategies[6, 7]. Unlike a treatment recorded as received, a prediction model may contribute to care at several stages before any treatment is delivered[1]. Prediction models therefore introduce an additional problem: unlike treatment received, individual-level model use may not be recorded in observational data, and even recorded use does not specify how clinicians respond to recommendations[1].

Model implementation involves several distinct stages. Clinicians may follow or modify a recommendation, and intended treatment decisions may not be executed. Consequently, the same model and threshold can produce different outcomes under different implementation processes. Population-level utilization, adherence, and execution rates may also conceal important differences in which patients receive or forgo treatment. Moreover, agreement between observed treatment and a reconstructed recommendation does not establish that the model influenced the clinician; the same treatment may have been chosen without model input. Thus, evaluating model impact requires well-defined implementation strategies and corresponding clinical outcome estimands[6, 7]. In particular, a comparison labelled simply as "model-guided care versus usual care" would leave unanswered which clinician responses and execution processes are included in the model-guided strategy. Observational records pose a separate issue: the care pathway documented in the data need not reveal the full implementation process.

We propose a framework that distinguishes model use, clinician decisions, and treatment execution; defines alternative target implementation strategies and their clinical consequences;

and characterizes their discrete and continuous comparisons. We also classify three observational data scenarios by the presence and observability of model use. By distinguishing target implementation strategies from the observed care pathways, the framework separates what an impact evaluation seeks to compare from what is recorded in routine care. The resulting estimands are relevant to impact studies regardless of whether the eventual evaluation is randomized or observational.

# 2 Conceptualizing Prediction Model Implementation

## 2.1 Notation and key variables

We consider a target population $\Omega$. For individual $i$, $M_i \in \{0,1\}$ indicates whether a model recommendation enters clinical decision making, distinguishing actual use from mere availability (Table 1). A fixed prediction model generates a predicted probability $\hat{p}_i$ and recommendation $R_i = I(\hat{p}_i \geq \tau)$ at a prespecified decision threshold $\tau$ [8]. The predicted probability is treated as a fixed output of the existing model and is used to determine $R_i$. It is not assumed to equal the counterfactual outcome risk under each implementation strategy, because the strategy may change treatment and hence the outcome distribution. The present framework considers a binary treatment decision, so $R_i, D_i, A_i \in \{0,1\}$, with 1 denoting recommendation, intention, or receipt of treatment, respectively. We distinguish the clinician's intended treatment decision $D_i$ from treatment received $A_i$ and denote the outcome by $Y_i$. Potentially overlapping covariate sets $C_i$ and $C_i'$ relate to implementation and treatment-outcome processes, respectively.

## 2.2 Sequential structure of prediction model implementation

The sequential implementation process distinguishes three stages (Figure 1): whether the recommendation enters clinical decision making; how a clinician responds; and whether the intended treatment is executed. We describe these relationships using potentially stochastic conditional mechanisms. The distinction between model availability and use is essential: installing or activating a model at a site does not imply that its output entered every individual decision. Even if the prediction can be reconstructed later, this does not reveal whether the clinician saw it.

**From model use to clinician decision.** The intended decision follows

$$D_i \sim \begin{cases} f_D^{(0)}(\cdot \mid C_i), & M_i = 0, \\ f_D^{(1)}(\cdot \mid R_i, C_i), & M_i = 1. \end{cases}$$

Here $f_D^{(0)}$ describes decision making without model input and $f_D^{(1)}$ when the clinician has considered the recommendation. If the predicted probability $p_i$ is also displayed and may influence decisions beyond the recommendation itself, $p_i$ can additionally be included in $f_D^{(1)}$. Model use does not imply adherence: clinicians may follow or modify $R_i$, and the same recommendation and covariates need not produce identical decisions. The mechanisms describe conditional patterns of decision making rather than fixed adherence proportions. For example, clinicians may deviate from recommendations for particular patient subgroups or because of information not included in the model. This heterogeneity matters when translating model use into a patient-level strategy. Accordingly, $f_D^{(1)}$ represents how decisions depend on the recommendation and clinical information, rather than a rule that clinicians must follow the recommendation. The corresponding no-model mechanism $f_D^{(0)}$ may use the same clinical information but cannot be influenced by a recommendation that was not consulted.

**From intended decision to treatment received.** Execution follows

$$A_i \sim f_A(\cdot \mid D_i, C_i).$$

The mechanism $f_A$ incorporates factors affecting delivery, such as resources, institutional restrictions, or patient preferences. Agreement $A_i = D_i$ indicates execution; disagreement identifies an execution gap. Therefore, $A_i \neq R_i$ alone cannot distinguish clinician modification from nonexecution. Conversely, agreement $A_i = R_i$ need not imply that a recommendation was followed: the clinician may have independently selected the same treatment. Observed treatment-recommendation concordance should not be equated with clinician adherence, which concerns $D_i$ rather than $A_i$.

**From treatment to clinical outcome.** We represent the outcome-generating process as

$$Y_i \sim f_Y(\cdot \mid A_i, C_i').$$

This framework focuses on model effects mediated through treatment received. For simplicity, $f_A$ does not additionally depend directly on model use; system-level or other pathways could be incorporated in extensions. For example, model deployment may alter ordering systems, monitoring intensity, or clinician behavior even when a particular recommendation is not used. These extensions would require additional variables or pathways beyond the present treatment-focused process. The mechanisms describe the implementation process, but a target strategy must specify which are retained or replaced (Section 3). The distinction is especially important when a strategy combines deterministic components, such as $D = R$, with stochastic components, such as treatment execution subject to delivery constraints.

### 2.3 Population-level implementation parameters

Three population-level summaries characterize different components of implementation:

$$\pi_M = P(M = 1 \mid \Omega), \qquad \alpha = P(D = R \mid M = 1, \Omega), \qquad \beta = P(A = D \mid \Omega).$$

They represent model utilization, recommendation–decision concordance among model users, termed adherence here, and execution of intended decisions, respectively (Supplementary Table S1). Because $\alpha$ is defined by $D = R$, it does not indicate whether the clinician would have made a different decision without model input. Moreover, $\alpha$ is defined only for implementation processes in which $P(M = 1) > 0$ and is not a parameter of $g_0$. These marginal rates do not uniquely define a strategy. For example, the same adherence rate can arise from overrides concentrated among high-risk or low-risk patients, with different outcome consequences. Complete target strategies must therefore specify patient-level decision and execution mechanisms rather than marginal implementation rates alone. Similarly, an unchanged execution mechanism $f_A$ may yield different overall execution rates when the distribution of intended decisions changes; the rate $\beta$ is a property of the resulting population-level process, not a complete description of that mechanism.

## 3 Defining the Values of Prediction Model Implementation

### 3.1 Target Implementation Strategies and Strategy-Specific Values

An implementation strategy $g$ specifies model use, the clinician decision mechanism, and how intended decisions become treatment received. Figure 2 distinguishes no-model care from four model-guided strategies forming a two-by-two structure: strict adherence versus clinician modification, and full versus constrained execution. These strategies are defined for the same target population and prediction rule, ensuring that differences in expected outcomes can be attributed to the specified implementation strategy rather than changes in the model itself. They describe target policies, not merely labels for naturally observed patient groups. The four model-guided strategies are generated systematically by crossing two clinician-response rules with two execution rules, rather than by assuming that implementation progresses through a single sequence of increasingly realistic stages.

**No-model care ($g_0$):** $M = 0$, $D \sim f_D^{(0)}(\cdot \mid C)$, and $A \sim f_A(\cdot \mid D, C)$. No-model care does not necessarily imply full execution. It coincides with usual care only when usual care excludes model input; after model implementation, observed usual care may include both model-informed and no-model pathways.

**Strict model-guided care ($g_1$):** $M = 1$, $D = R$, and $A = D$; recommendations are followed and executed.

**Clinician-modified care ($g_2$):** $M = 1$, $D \sim f_D^{(1)}(\cdot \mid R, C)$, and $A = D$; clinicians may modify recommendations, but decisions are executed.

**Strict guidance with execution constraints ($g_3$):** $M = 1$, $D = R$, and $A \sim f_A(\cdot \mid D, C)$.

**Clinician modification with execution constraints ($g_4$):** $M = 1$, $D \sim f_D^{(1)}(\cdot \mid R, C)$, and $A \sim f_A(\cdot \mid D, C)$.

Strategies $g_1$ and $g_2$ serve as idealized complete-execution benchmarks. Here, full execution means $A = D$ for an appropriately defined intended decision. When patient preferences, contraindications, or other considerations are intended to remain part of the strategy, they should be incorporated into the definition of $D$ or retained in the treatment-execution mechanism.

The mechanisms retained under a strategy must be specified sufficiently to characterize the patient-level process; they need not equal those operating in a particular dataset. For example, "clinician modification" requires more than allowing an unspecified fraction of recommendations to be overridden: the strategy must describe how intended treatment decisions depend on $R$ and $C$. Similarly, constrained execution requires a rule for how received treatment depends on the intended decision and relevant circumstances. In contrast, $g_1$ defines decisions and execution deterministically and requires no separate adherence or execution-rate parameter. The same execution mechanism can yield different marginal execution rates when the distributions of intended decisions differ across strategies. In particular, $g_0, g_3$ and $g_4$ retain $f_A$, while $g_1$ and $g_2$ impose full execution. There is no presumed ranking of the strategies by clinical desirability: clinician modification can correct unsuitable model recommendations, while execution barriers may have consequences that depend on the proposed treatment.

Let $Y_i^g$ denote the potential outcome for individual $i$ under $g$. The *implementation value* is the expected outcome for the target population:

$$V(g) = E_{\Omega}[Y^g].$$

Here *value* does not necessarily mean utility; which direction is favorable depends on the outcome definition. For mortality or another adverse outcome, a smaller $V(g)$ may be preferable; for a favorable outcome, a larger value may be preferred. The framework therefore defines a comparison without presupposing a favorable effect of introducing or strictly following model guidance. The same notation can accommodate scalar outcomes other than binary events, provided the outcome and follow-up period are prespecified. For a time-to-event endpoint, $Y^g$ should denote a defined scalar summary, such as event status by a fixed horizon or restricted survival time. Because every $V(g)$ refers to the same $\Omega$, the contrasts compare target policies in a common population.

### 3.2 Discrete Contrasts Between Implementation Strategies

For strategies $g_a$ and $g_b$, define the implementation effect as

$$\Delta(g_a, g_b) = V(g_a) - V(g_b).$$

Table 2 presents six prespecified contrasts. Two compare model-guided strategies with no-model care, and four compare pairs of model-guided strategies that differ in one implementation component. The latter contrasts isolate clinician modification under a fixed execution rule or execution constraints under a fixed clinician-decision rule. Specifically, $V(g_1) - V(g_0)$ compares a strategy imposing both strict adherence and full execution with no-model care, whereas $V(g_4) - V(g_0)$ compares model-informed care allowing clinician modification and execution constraints with no-model care. Thus, $V(g_1) - V(g_0)$ should not be interpreted as isolating the effect of model guidance while holding the execution mechanism fixed. The contrasts $V(g_2) - V(g_1)$ and $V(g_4) - V(g_3)$ isolate clinician modification under different execution conditions; the other two contrasts isolate execution constraints under different clinician decision mechanisms. The resulting four model-guided strategies also permit a factorial interaction contrast, although its interpretation depends on their common mechanism definitions. One such contrast is $[V(g_4) - V(g_3)] - [V(g_2) - V(g_1)]$: it asks whether the expected-outcome difference associated with clinician modification changes when execution is constrained rather than complete.

### 3.3 Continuous Implementation Effects and the Value Surface

Adherence and execution can vary continuously within a specified family of patient-level implementation mechanisms. A change in the overall adherence rate must be associated with a prespecified way in which individual clinicians alter their decisions. For instance, increasing adherence among patients near a treatment threshold is not equivalent to increasing it among patients far above that threshold, even if both changes yield the same marginal rate. Thus, specifying $g_{\alpha,\beta}$ requires a rule for which decisions or executions change as each parameter changes; the rates alone do not define such a rule. With $M = 1$, let $g_{\alpha,\beta}$ be the strategy whose specified conditional mechanisms generate rates $\alpha$ and $\beta$ in $\Omega$. Define

$$V(\alpha, \beta) \equiv V(g_{\alpha,\beta}) = E_{\Omega}[Y^{g_{\alpha,\beta}}].$$

Varying $\alpha$ and $\beta$ over feasible values defines an **implementation value surface**. The feasible domain depends on the chosen mechanism family: not every pair of marginal rates must be attainable, and smooth changes in the rates need not always produce a differentiable surface. The following local summaries apply where the relevant variation and derivatives exist. Where differentiable, $\partial V / \partial \alpha$ and $\partial V / \partial \beta$ describe local sensitivity to changes in adherence and execution generated by that mechanism family. Their signs do not imply that greater adherence or execution always improves outcomes. Rather, the partial derivatives quantify local changes in the expected outcome under the specified mechanism family, holding the other implementation dimension fixed. These quantities may help describe the relative sensitivity of clinical consequences to the two dimensions; determining implementation priorities also requires costs and feasibility.

Contours represent combinations producing the same expected outcome. The break-even contour $V(\alpha, \beta) = V(g_0)$, when present, identifies combinations matching no-model care. The mixed derivative $\partial^2 V/(\partial\alpha\ \partial\beta)$ describes interaction between the two implementation dimensions. In particular, it describes whether the outcome sensitivity to changes in adherence depends on the execution level. An interaction expressed on the expected-outcome scale should not automatically be interpreted as a beneficial synergy.

Because identical marginal rates can arise from different patient-level processes, $V(\alpha, \beta)$ is **not universal**: it depends on the specified mechanism family. Discrete strategies correspond to points on a common surface only if they belong to the same family (Supplementary Section S2). The no-model strategy $g_0$ is an external reference rather than a point on the model-guided surface because $M = 0$ rather than $M = 1$. Consequently, the break-even contour compares model-guided strategies on the surface against a separately defined no-model reference; it should not be interpreted as a contour of zero adherence.

## 4 Observational Data Scenarios for Evaluating Prediction Model Implementation

The implementation information available in observational data depends on whether the model was introduced into care and whether individual-level use $M_i$ was recorded. We distinguish three scenarios (Table 3). The scenarios do not imply that the model was used uniformly across the healthcare system; instead, they separate model availability or deployment from whether an individual recommendation actually informed care. In each, $\hat{p}_i$ and $R_i$ may be reconstructed if the model, predictors, and threshold are available; retrospective reconstruction does not establish actual use. Routine records may document $A_i$ without separately recording $D_i$, obscuring clinician modification and execution gaps. Likewise, the mere presence of a risk score in a database does not establish that it was shown to a clinician before a decision was made. Thus, missing information about the intermediate implementation process is a general concern across scenarios, even when model-use logs exist.

### 4.1 Scenario I: Prediction model not implemented

The model has not entered routine clinical decision making, so $M_i = 0$ for all individuals. Observed care follows the no-model pathway. Although recommendations can be reconstructed retrospectively, no model-informed care pathway is present in these records. This is a description of care that actually occurred, not of the hypothetical recommendations that a model-guided policy might have generated for the same patients.

### 4.2 Scenario II: Prediction model implemented with observed model use

The model has entered clinical practice, and $M_i$ is recorded. Care with model input ($M_i = 1$) can therefore be distinguished from care without it ($M_i = 0$), when both occur. Model input need not imply adherence, nor does an intended decision necessarily equal treatment received. A record that the clinician reviewed the recommendation before finalizing the treatment decision supports classification as $M_i = 1$, even when the original plan is retained. If the log records only that the recommendation appeared on screen, individual model use remains uncertain.

### 4.3 Scenario III: Prediction model implemented without observed model use

The model is implemented, but $M_i$ is unrecorded. The data may contain an unlabeled mixture of model-informed and no-model care. Agreement between reconstructed recommendation $R_i$ and actual treatment $A_i$ cannot establish use, because the same treatment may have been chosen without the model. Conversely, disagreement cannot establish non-use: clinicians may have modified recommendations or intended treatment may not have been executed. The scenario need not contain both pathways in nonzero proportions; the defining feature is that their patient-level attribution is unrecorded.

These scenarios characterize observed data structures, not different target strategies. They establish what information about implementation is available while leaving the corresponding strategy-specific clinical consequences conceptually unchanged. Distinguishing strategies from observed data structures prevents an observed group or mixture from being mistaken for a fully specified target policy.

More specifically, Scenario I contains no observations of the model-informed decision mechanism $f_D^{(1)}$; evaluation of $g_2$ and $g_4$ therefore requires this mechanism to be specified using external evidence or assumptions. In Scenario II, recording $M$ distinguishes model-informed from no-model pathways, but $D$ must also be observed to separate clinician modification from treatment execution. In Scenario III, unrecorded $M$ additionally prevents direct attribution of observed pathways to $f_D^{(0)}$ or $f_D^{(1)}$. These observations concern data requirements; identification of $V(g)$ still requires additional causal assumptions.

## 5 Discussion

We developed a framework that separates prediction model use, clinician decision making, and treatment execution, and translates this process into target implementation strategies and clinical outcome estimands. Three observational scenarios additionally characterize how much of the process is recorded in routine data. The resulting distinction between target strategies and observed care provides a common language for describing the questions that prediction model impact studies actually address.

Prior work has emphasized evaluating prediction models beyond discrimination and calibration[1, 4]. Our framework highlights why an indicator of availability or use is insufficient: clinicians may modify recommendations, and intended treatments may not be delivered. Even with an unchanged model and decision threshold, different implementation processes may lead to different outcomes. Interpreting an impact comparison therefore requires specifying how care would be delivered under each strategy. For example, in a decision-analytic study of the HEART score, Jenniskens et al.[9] showed that projected outcomes and costs varied with how often clinicians followed the recommendations and whether they used additional clinical information when departing from them. Our framework also considers what happens after the clinician has made a decision. Even when a clinician follows a recommendation and prescribes treatment, the patient may never receive it. This gap between recommended and received treatment may also matter when interpreting a decision curve analysis, in which the conventional model-based curve represents the net benefit of following a threshold-based treatment rule[2]. Future work could construct a decision curve for each of the four model-guided strategies and use usual care as a common comparator to examine whether clinician modification or incomplete treatment execution changes the range of thresholds over which model-guided care offers greater net benefit.

The five strategies distinguish no-model care from combinations of strict or modified decisions and full or constrained execution. Their contrasts address different clinical questions and cannot generally be reduced to changes in overall adherence or execution rates. The value surface extends these comparisons to continuous variation within specified patient-level mechanisms. More adherence or execution is not automatically beneficial; consequences depend on who is treated, treatment effects, and outcome direction. Even a small difference in marginal adherence can have substantial consequences if concentrated in patients for whom the treatment decision is consequential, while a larger change elsewhere may have little effect.

This distinction matters particularly for observational evaluations. Electronic health records document many treatments and outcomes but may not record model use or intended clinician decisions[10]. Agreement between reconstructed model recommendations and received treatment cannot establish model use; disagreement may arise from clinician modification or an execution gap. Recording model-use logs, intended decisions, treatment delivery, and reasons for deviations would improve descriptions of implementation. In both prospective impact studies and routine-data research, these records would help distinguish whether a recommendation failed to influence the intended decision or whether the intended action was not delivered. Reporting the implementation strategy explicitly would also improve comparisons across studies using the same prediction model but different clinical workflows. Clear target strategies can also inform subsequent observational evaluations, including target trial emulation[6, 7]. Before selecting an estimator, investigators should specify the target strategy and determine whether the recorded variables adequately describe its relevant implementation stages.

The framework is conceptual and focuses on individual-level model use and treatment-mediated clinical consequences. It does not explicitly represent system-level changes in workflows or other effects of model availability, and its execution mechanism is assumed not to depend directly on model use. The five strategies are illustrative rather than exhaustive. They assume a fixed model and treatment decision rule; implementation involving repeated predictions, multiple treatment options, partial model uptake, or changing thresholds would require further strategy definitions. The three observational scenarios classify available information but do not establish identifiability or specify estimators. Additional assumptions, including transportability when mechanisms are taken from external settings, are required for empirical evaluation[11].

In conclusion, prediction model impact cannot be adequately defined by model use alone. Explicit implementation strategies and strategy-specific outcomes clarify what impact evaluations should compare and how these questions relate to observational data.

## Ethics approval

Ethics approval was not required because this article presents a conceptual methodological framework and did not involve human participants or identifiable individual-level data.

## Acknowledgements

Not applicable.

## Authors' contributions

WJF and WSF conceived the study and developed the initial conceptual framework. WJF and YYZ refined the conceptual and methodological structure of the framework and interpreted its implications for observational evaluation of prediction-model implementation. WJF and YYZ conducted the literature review. YYZ drafted the manuscript. WJF and WSF critically revised the manuscript for important intellectual content. All authors contributed to subsequent revisions and approved the final version of the manuscript. WJF and WSF are the guarantors for the paper.

## Funding

The paper is funded by the National Natural Science Foundation of China (72342015), and the National Key Research and Development Program of China (2023YFC2507200, 2026ZD0556201).

## Conflict of interest

None declared.

## Use of artificial intelligence tools

ChatGPT and DeepSeek were used to assist with language editing and the presentation of methodological concepts. All AI-assisted content was critically reviewed and revised by the authors, who take full responsibility for the accuracy and integrity of the final manuscript.

## Tables

**Table 1. Definitions of variables in the prediction model implementation framework**

| Feature | Definition | Interpretation |
|---|---|---|
| Target population $\Omega$ | Population for the impact evaluation | All strategy values refer to this population. |
| Model use $M_i$ | Whether recommendation enters decision making | Distinct from system-level availability. |
| Prediction $\hat{p}_i$ | Predicted outcome probability | Reconstructable if the model and required predictor data are available, regardless of actual use. |
| Recommendation $R_i$ | $I(\hat{p}_i \geq \tau)$ | Treatment recommendation based on fixed threshold $\tau$. |
| Intended decision $D_i$ | Clinician's treatment intention | May differ from $R_i$. |
| Treatment received $A_i$ | Actual treatment delivered | May differ from $D_i$. |
| Outcome $Y_i$ | Clinical outcome of interest | A prespecified scalar clinical outcome over a defined follow-up period, such as event status by a fixed horizon, a continuous outcome, restricted survival time, or cost. |
| Covariates $C_i, C_i'$ | Implementation and outcome-related factors | Sets may overlap. |

**Table 2. Discrete contrasts between implementation strategies**

| Contrast | Expression | Component(s) changed | Interpretation |
|---|---|---|---|
| Strict model-guided care | $V(g_1) - V(g_0)$ | Model use, decision, execution | Strict model guidance versus no-model care. |
| Model-guided care with modification and constraints | $V(g_4) - V(g_0)$ | Model use and decision; same execution mechanism | Model-informed care with clinician modification and execution constraints versus no-model care. |
| Clinician modification under full execution | $V(g_2) - V(g_1)$ | Decision mechanism | Allow clinician modification while retaining full execution. |
| Execution constraints under strict adherence | $V(g_3) - V(g_1)$ | Execution mechanism | Replace full execution with constrained execution while retaining strict adherence. |
| Execution constraints under clinician modification | $V(g_4) - V(g_2)$ | Execution mechanism | Replace full execution with constrained execution while retaining the clinician-modification mechanism. |
| Clinician modification under execution constraints | $V(g_4) - V(g_3)$ | Decision mechanism | Allow clinician modification while retaining the constrained-execution mechanism. |

**Table 3. Observability of key variables across observational scenarios**

| **Feature** | **Scenario I: Model not implemented** | **Scenario II: Model implemented, use observed** | **Scenario III: Model implemented, use unobserved** |
|---|---|---|---|
| Model use $M$ | Structurally 0 | Recorded | Not recorded |
| Prediction and recommendation | Reconstructable | Reconstructable | Reconstructable |
| Intended decision $D$ | Often not recorded separately | Often not recorded separately | Often not recorded separately |
| Treatment received $A$; outcome $Y$ | Typically recorded | Typically recorded | Typically recorded |
| Observed care pathway | No-model only | Pathways distinguishable when present | Potentially unlabeled mixture |

*Note:* Reconstruction presupposes availability of the model, predictors, and decision rule; actual variable completeness depends on the data source.

# Figures

## Figure 1: The Flowchart of the Sequential Prediction Model Implementation Pathway

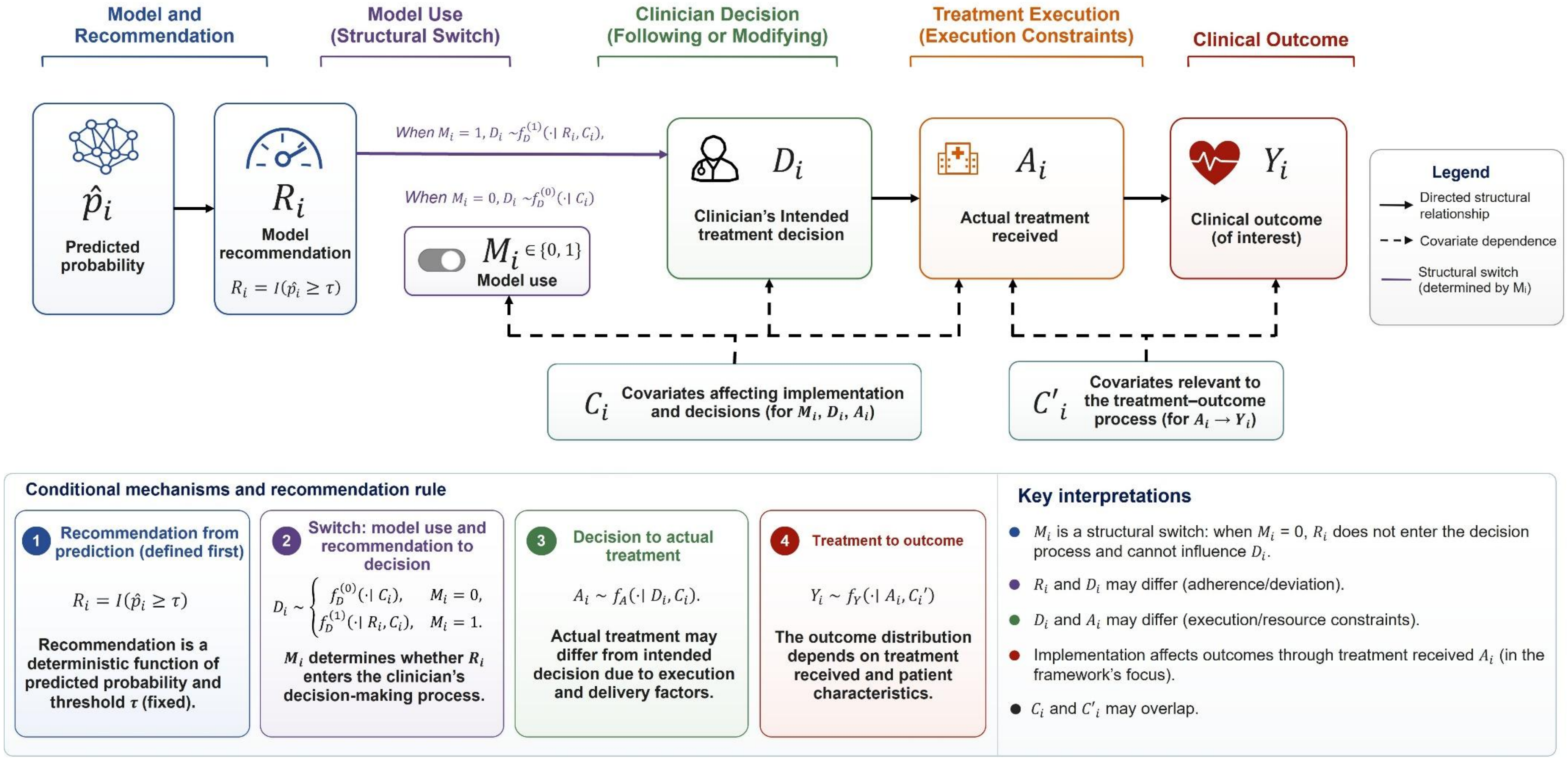

**Figure 2. Target Implementation Strategies for Model-Guided Care and Their Corresponding Implementation Values**

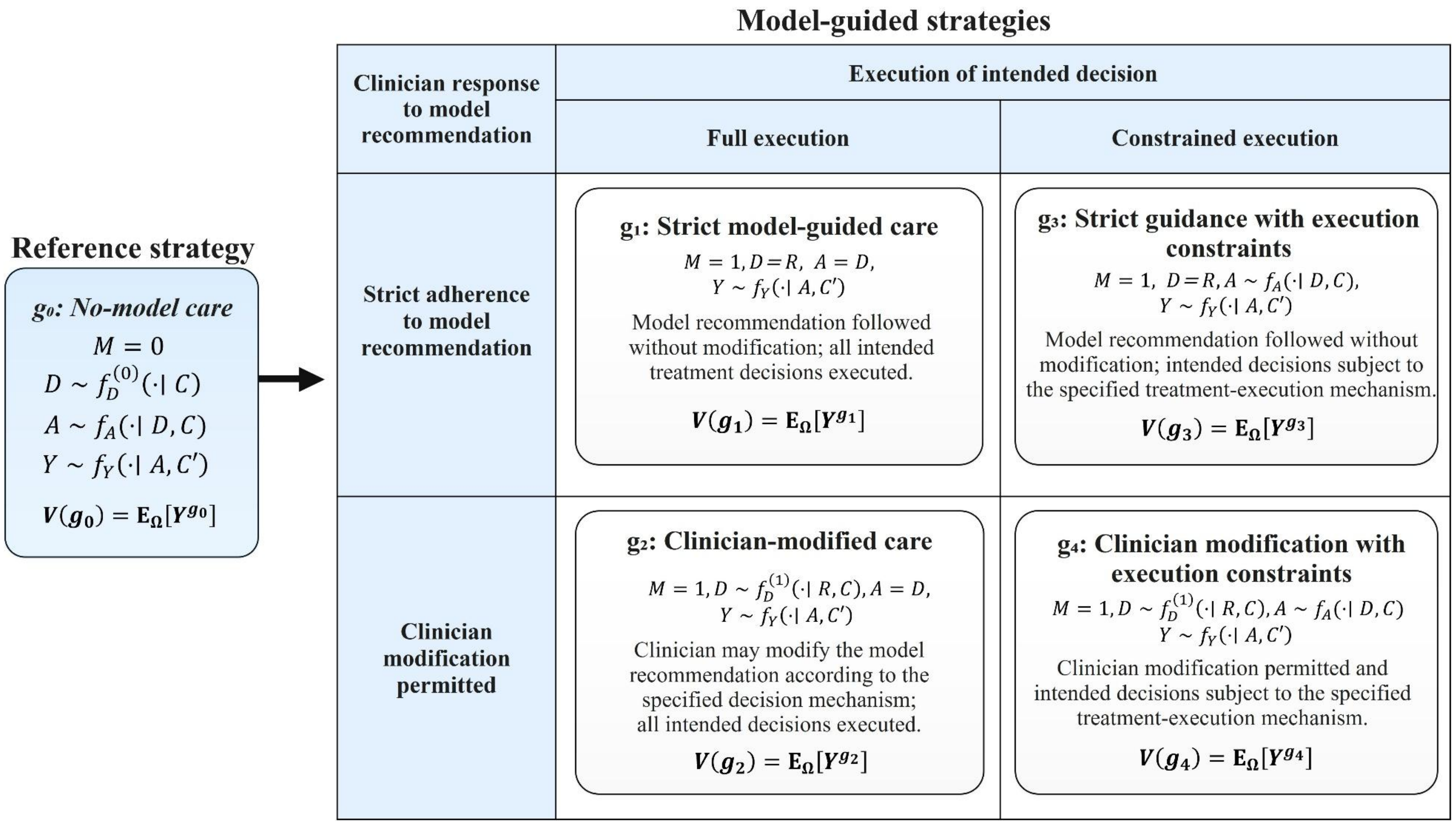

**Notes:** $g_0$ represents no-model care (reference strategy). No-model care does not imply full execution. When the prediction model is used ($M = 1$), strategies differ in how clinicians respond to model recommendations and whether intended treatment decisions are fully executed or governed by the specified treatment-execution mechanism, yielding four model-implementation strategies $g_1 - g_4$ and their corresponding implementation values $V(g)$. These strategies are not ranked by clinical desirability.

# Supplementary Materials

## Supplementary Table S1. Population-level parameters describing prediction model implementation

| Parameter | Definition | Interpretation |
|---|---|---|
| Model utilization rate $\pi_M$ | $P(M = 1 \mid \Omega)$ | Proportion of individuals in the target population for whom a model recommendation enters clinical decision making. This differs from system-level model availability. |
| Clinician adherence rate $\alpha$ | $P(D = R \mid M = 1, \Omega)$ | Agreement between the clinician's **intended** treatment decision and the model recommendation among model users. It is not the same as treatment–recommendation concordance $P(A = R \mid M = 1, \Omega)$. |
| Treatment execution rate $\beta$ | $P(A = D \mid \Omega)$ | Agreement between treatment received and the intended decision, irrespective of model use. Different strategies can yield different marginal execution rates even if they share a conditional execution mechanism. |

**Note.** $\Omega$ denotes the target population; $M$ model use; $R$ the model recommendation; $D$ the clinician's intended treatment decision; and $A$ treatment received. These rates summarize the implementation process but do not, by themselves, define a patient-level implementation strategy. The same marginal rate can reflect different conditional patterns across patient subgroups and thereby lead to different clinical consequences.

# Supplementary Section S2. Continuous implementation effects

## S2.1. Defining a family of patient-level implementation mechanisms

The value surface in Section 3.3 is defined for a prespecified family of model-guided strategies, not for arbitrary combinations of marginal adherence and execution rates. Let

$$\mathcal{G} = \{g_{\alpha,\beta}: (\alpha, \beta) \in \mathcal{D}\},$$

where $\mathcal{D} \subseteq [0,1]^2$ contains the pairs attainable within that family. Each $g_{\alpha,\beta}$ sets $M = 1$ and specifies conditional decision and execution mechanisms. Write $q_{D,\alpha}(d \mid r, c)$ for the probability of intended decision $d$ under recommendation $r$ and covariates $c$, and $q_{A,\alpha,\beta}(a \mid d, c)$ for the probability of treatment received $a$ given intended decision $d$ and covariates $c$. The indexing of the latter mechanism allows the execution rule, if necessary, to be calibrated when the distribution of intended decisions changes.

For binary decisions and treatments, the rates generated in the target population must satisfy

$$\alpha = E_{\Omega}\left[\sum_{d=0}^{1} I\,(d = R) q_{D,\alpha}(d \mid R, C)\right],$$

$$\beta = E_{\Omega}\left[\sum_{d=0}^{1}\sum_{a=0}^{1} I\,(a = d)\, q_{A,\alpha,\beta}(a \mid d, C)\, q_{D,\alpha}(d \mid R, C)\right].$$

These equalities clarify that $\alpha$ and $\beta$ are **generated by** the conditional mechanisms, rather than uniquely specifying them. In particular, the marginal execution rate can depend on the clinician decision distribution. With $Y^{g_{\alpha,\beta}}$ denoting the potential clinical outcome under the complete strategy, the implementation value surface is

$$V(\alpha, \beta) \equiv V\left(g_{\alpha,\beta}\right) = E_{\Omega}[Y^{g_{\alpha,\beta}}].$$

The specification concerns target strategies and their potential outcomes; it does not by itself establish identification from a particular observational dataset.

## S2.2. An illustrative stochastic implementation mechanism

A simple mathematical construction demonstrates how a complete mechanism family can be indexed by adherence and execution. Suppose $R, D, A \in \{0,1\}$, model use is fixed at $M = 1$, and the following two conditional rules apply:

$$D = \begin{cases} R, & \text{with probability } \alpha, \\ 1 - R, & \text{with probability } 1 - \alpha, \end{cases}$$

$$A = \begin{cases} D, & \text{with probability } \beta, \\ 1 - D, & \text{with probability } 1 - \beta. \end{cases}$$

The random draws are specified to follow these probabilities conditionally at each stage. This family produces $P(D = R \mid M = 1, \Omega) = \alpha$ and $P(A = D \mid \Omega) = \beta$, with

$$P(A = R \mid \Omega, M = 1) = \alpha\beta + (1 - \alpha)(1 - \beta).$$

The latter expression shows why agreement between the observed treatment and the recommendation is not equivalent to clinician adherence. The rule allowing $A = 1$ after an intended decision $D = 0$ is an artificial binary switch used only to illustrate the mathematics; actual nonexecution may be asymmetric and patient-dependent. More realistic mechanisms should specify such asymmetries explicitly.

**Why marginal rates are insufficient.** Consider two equally sized patient subgroups, H and L, in which $R = 1$ for everyone and intended decisions are always executed ($\beta = 1$). Under strategy $g^{(a)}$, adherence probabilities are $0.9$ in H and $0.5$ in L; under $g^{(b)}$, they are $0.5$ in H and $0.9$ in L. Within each subgroup, adherence is randomly assigned according to these probabilities, independently of individuals' potential outcomes under treatment and no treatment. Both have the same marginal adherence $\alpha = 0.7$, but they allocate adherence differently across subgroups. These strategies illustrate why marginal rates alone do not specify a strategy and why a mechanism family must be specified before defining $V(\alpha, \beta)$. Suppose $Y$ is an adverse outcome, let $\mu_{a,k} = E(Y^a \mid k)$ denote the expected outcome under treatment $a$ in subgroup $k$, and assume $\mu_{1,H} - \mu_{0,H} = -0.20$ and $\mu_{1,L} - \mu_{0,L} = 0$. Then

$$V\left(g^{(a)}\right) - V\left(g^{(b)}\right) = \frac{1}{2} \times (0.9 - 0.5) \times (-0.20) + \frac{1}{2} \times (0.5 - 0.9) \times (0) = -0.04.$$

Thus, identical $\alpha$ and $\beta$ do not guarantee identical implementation values. The numerical values are illustrative and not empirical estimates.

## S2.3. Marginal effects and break-even contours

Where $V(\alpha, \beta)$ is differentiable on the interior of $\mathcal{D}$, the partial derivatives

$$\frac{\partial V(\alpha,\beta)}{\partial \alpha} \quad \text{and} \quad \frac{\partial V(\alpha,\beta)}{\partial \beta}$$

describe local changes in expected outcome associated with the **specified** changes in the patient-level mechanisms. Their magnitudes depend on which patients experience changes in clinician decisions or treatment execution. Neither derivative has a universally favorable sign, and comparisons across different mechanism families require care.

For a target outcome value $v$, a contour is the level set

$$\mathcal{C}_v = \{(\alpha,\beta) \in \mathcal{D}: V(\alpha,\beta) = v\}.$$

A break-even contour relative to no-model care is $\mathcal{C}_{V(g_0)}$, if nonempty. Whether values above or below this contour are preferred depends on the outcome definition. A contour describes expected clinical consequences; it does not imply that all corresponding implementation strategies are feasible or estimable from an available dataset.

# S2.4. Discrete and continuous interactions

The mixed partial derivative

$$\frac{\partial^2 V(\alpha,\beta)}{\partial \alpha\, \partial \beta}$$

measures how the local sensitivity to adherence varies with execution (and vice versa), on the expected-outcome scale. A corresponding discrete interaction can be defined using four points in a **single** mechanism family. For $\alpha_L < \alpha_H$ and $\beta_L < \beta_H$, define

$$\Delta_{\text{int}} = [V(\alpha_L,\beta_L) - V(\alpha_H,\beta_L)] - [V(\alpha_L,\beta_H) - V(\alpha_H,\beta_H)].$$

If the rectangle $[\alpha_L,\alpha_H] \times [\beta_L,\beta_H]$ lies entirely within the feasible domain $\mathcal{D}$, and $V$ has a continuous mixed partial derivative on an open neighborhood of this rectangle, then

$$\Delta_{\text{int}} = \int_{\alpha_L}^{\alpha_H} \int_{\beta_L}^{\beta_H} \frac{\partial^2 V(\alpha,\beta)}{\partial \alpha\, \partial \beta}\, d\beta\, d\alpha.$$

When $g_1, g_2, g_3, g_4$ coincide respectively with the four corners $(\alpha_H,\beta_H)$, $(\alpha_L,\beta_H)$, $(\alpha_H,\beta_L)$, and $(\alpha_L,\beta_L)$ of this same mechanism family, the discrete contrast also equals $[V(g_4) - V(g_3)] - [V(g_2) - V(g_1)]$. **This correspondence is conditional**, not automatic: the four strategies defined in the main text need not belong to a common continuous mechanism family. Moreover, the no-model strategy $g_0$, for which $M = 0$, lies outside the model-guided surface with $M = 1$.